Neural computation from first principles: Using the maximum entropy method to obtain an optimal bits-per-joule neuron

William B Levy[1], Toby Berger[2], Mustafa Sungkar[2]

[1] Dept.'s of Neurosurgery and of Psychology, School of Medicine, University of Virginia, Charlottesville, VA 22908
[2] Dept. Electrical and Computer Engineering, University of Virginia, Charlottesville, VA 22903

**ABSTRACT**
Optimization results are one method for understanding neural computation from Nature's perspective and for defining the physical limits on neuron-like engineering. Earlier work looks at individual properties or performance criteria and occasionally a combination of two, such as energy and information. Here, as the optimization method, we make use of Jaynes' maximum entropy method and point out some of the different types of constraints, possibly dimensionally distinct, the method can combine. A neuron, as a computational device, is assumed to estimate a scalar latent variable and to encode this estimate with an interpulse-interval. Arising from each such constraint set, the inference method identifies a likelihood-function and a sufficient statistic for the estimate. This likelihood is a first-hitting time distribution in the exponential family. Particular constraint sets are identified that, from an optimal inference perspective, align with earlier neurocomputational models. Interactions between constraints, mediated through the inferred likelihood, restrict constraint-set parameterizations, e.g., the energy-budget limits action potential threshold which limits estimation performance. Such linkages are, for biologists, experimental predictions arising from the method. In addition to the likelihood, which is a conditional distribution of the interpulse interval given the variable being estimated, at least one type of constraint set restricts the two marginal distributions. In this case, a bits/joule statement arises using Lindley's interpretation of a Bayesian experiment to get bits per pulse-out.

I. INTRODUCTION

First principle/optimization approaches for understanding neural communication and computation at the level of a single neuron, predicting firing rates or quantitative connectivity or both, are rare but possible with careful selection of constraints and variables as combined with an appropriate optimization method [1-8].

Energy-efficient sensing, computation, and communication are among these optimization schemes. (Without contradiction, energy can be measured in various units: moles of $O_2$, glucose, ATP, or joules (J) *inter alia*.) The biological motivation for energy-based constraints is strong since (i) energy-use is a common currency for costs, applicable across every physiological level — subcellular, cellular, organ-level, and even to an organism's behavior and (ii) energy-use is generically relatable to biological fitness and reproductive success [9,10]. In other words, quantified, energy-constrained optimizations use Darwin's paradigm (microscopic optimization through natural selection) to create a restricted range of mathematical models for the physiological problem being addressed. From another direction, there are strong arguments from physics [11] that repeated sensing and memory storage are ultimately energy-limited. Thus, when successfully executed, the energy-constrained optimization approach promises both a way to understand brain-based computation and a way to specify the ultimate limits on engineered neural computation. Here we present a methodology that allows one to combine other constraints, in addition to energy-use, to define optimal computation.

It is generally agreed that a neuron performs a computation. However in any specific situation, the exact nature of such a computation is a matter of conjecture. Here we conjecture that a neuron estimates the value of a scalar, latent variable, and we derive a family of optimal-inference probability densities for such an estimate. In other words, instead of merely viewing a neuron as a communication device [1,2,4,5,12], which optimizes mutual information just like a communication channel, the hypothesis here is that a neuron's computation (estimate) corresponds to the maximum entropy distribution constrained by a set of expectation constraints that define and/or limit the neuron's operation.

Technically, the starting point is Jaynes' maximum entropy method (MEM) [13,14, and see REMARKS A for a variety of motivations for using the MEM] and its extension, the minimum relative entropy method [15]. Such methods can incorporate a multiplicity of constraints that are in the form of expectations to infer a probability distribution. The probability distributions to be inferred are interpulse (interspike) interval distributions. One such constraint is average energy-use. A second constraint arises by assuming that a neuron (or neural-like element) estimates the value of a scalar latent variable with each of its output pulses where each pulse-out is an interpulse interval (IPI) coding of this estimate. We then set our sights on constraint sets and a neuron in which such an IPI is a sufficient statistic for the value of the latent variable. Motivating this goal are the desirable, even optimal, characteristics of a sufficient statistic: it carries all the information of sampling; it can be, and will be here, the best unbiased minimum variance estimate; and it achieves the Cramer-Rao bound on estimation. Along the way, we are also able to relate increased synaptic energy-use to the reduction of the mean square error (MSE) of the estimation.

II. PRIMARY ASSUMPTIONS, SOME NOTATION, AND BRIEF SUMMARY OF RESULTS

As in [5,12], the critical pair of assumptions here concern the input and output variables, $\{\Lambda_j(k), T_j(k)\} := \{\Lambda, T\}$ with realizations written as $\{\lambda, t\}$, where, just in this section, $k$ indicates the $k^{th}$ IPI of a sequence of IPIs. That is, $t$ is the duration of the $k^{th}$ IPI, which ranges over the positive reals. The initial range of $\Lambda$ is similiar, but later, further restrictions arise from the calculations.

*A0*. A neuron, call it *j*, is constructed to estimate (i.e., compute) a particular non-negative, scalar, latent (hidden) RV, $\Lambda_j(k)$,. That is, the net synaptic excitation of *j* at any time is a function of $\lambda$'s value; on average net synaptic excitation is monotonically increasing in $\lambda$. More abstractly, $\lambda$ is the weighted sum of the inputs to *j*. This neuron sporadically communicates an estimate of the latent variable, $\hat{\lambda} := \hat{\lambda}_j(t(k))$ via a single pulse-out. This pulse is of fixed amplitude and fixed shape regardless of the IPI and when this pulse reaches a recipient of *j*'s output, the pulse is, more or less, instantaneously decoded. The shape and amplitude of *j*'s pulse is dictated by physical aspects of its axon, a dispersive transmission line. The neuron has an assigned, on-average, energy budget, $\mathcal{E}$. Any assumed expectation is finite, and those that are not are ruled out.

*Corollary to A0.* Assumptions within *A0* (fixed shape and amplitude pulses and instantaneous decoding) imply that the only information available to a recipient of *j* are the pulse-arrival-times, or equivalently, the IPIs, assuming an available timer; moreover, these IPIs cannot be used for block-coding that uses delays. Thus, the pulsewise, output random variable (RV), up to equivalence, must be the IPI, *T*.

*A1*. An IPI, *t*, is, on average, a monotonically decreasing function of the randomly generated latent value λ.

While not needed for the results in section III, it will help some readers get through it by being a little more concrete, as we must be for the results in section IV. First concerning the overall average rate of synaptic arrivals during one IPI, the random process Λ is the sum of the mean rates of the individual point processes governing each synaptic activation of each input to *j*. Second, the IPI is a first hitting-time (FHT). Thus, because *T=t* is a FHT, the probability distribution $p(t|\lambda)$ is a FHT distribution conditional on the latent variable that controls the net synaptic excitation.

*A2*. Assume that any input value, $\Lambda_j(k) = \lambda$, is stationary throughout *j*'s IPI and assume that there is Markovian conditional independence for successive IPIs with the necessary and locally available conditioning information stored by each of *j*'s recipients.

The assumptions in *A0* and *A1* seem fundamental to all the important results presented here. The assumptions in *A2* simplify the mathematical developments and will be supplanted in future developments with a less restrictive statement.

Brief summary of the main results:
(*a*) Three distinct types of constraints on neural computation are identified, and their forms are specified so that (i) $p(t|\lambda)$ is in the exponential family and (ii) the sufficient statistic of this density is *t* and/or an invertible function of *t*.
(*b*) Each IPI value, *t*, is equivalent to an estimate of λ, and the MSE of this estimate is a decreasing monotonic function of synaptic energy-use.
(*c*) There exists at least one constraint set and neural model that, beyond the implied generalized inverse Gaussian distribution of $p(t|\lambda)$, also implies the marginal distributional forms of $p(\lambda)$ and $p(t)$; this marginal distribution of *t* is not in the exponential family.
(*d*) The combination of (c), Bayes theorem, and the original energy-constraints implies *j*'s computationally generated bits/J. Thus *j* is implicitly computing a Bayesian inference as well as computing an estimation.

## III. OPTIMIZING A NEURON'S COMPUTATION

We will not explain here why maximizing the entropy leads to optimal inference of a probability distribution but if unfamiliar with this idea see REMARKS A and the references there cited. We use Theorem 11.1.1 in Cover and Thomas [16] to infer $p(t|\lambda)$. This theorem yields the unique distributional form, a continuous density, when using the MEM given a finite set of finitely valued expectations, i.e., these expectations are the constraints for a continuous RV. The proof of the theorem becomes a Lagrange multiplier problem, and its implementation the MEM that states and applies a constraint set ; here we take the functional derivative $\frac{\partial \cdot}{\partial p(t|\lambda)}$. Technically, the resulting terms in the exponent must be unitless which is accomplished either through the Lagrange multipliers, the subscripted $\alpha$'s appearing below, or as we sometimes prefer, the constants within the expectations (lower case *c* that will later by subscripted in various cases). Also from the axiomatic approach of Shore and Johnson (15), one must be careful that the constraint set is not self-contradictory or redundant (i.e., one particular expectation constraint is derivable from the others; see REMARKS B). With the above theorem in place, there is a simple corollary that leads us toward the sought after sufficient statistic. This corollary identifies the constraint forms that allow such a sufficient statistic.

*MEM Corollary*. For any $\Lambda=\lambda$ of the conditional probability $p(t|\lambda)$, given are the constraint functions and their expectations where the expectations are obtained by integration over the entire fixed range of *T*. These functions and associated expectations are assumed to be of three basic types: (A) functions $\tilde{g}_{A,h}(\lambda,t)$ that are limited to factorable functions and one special case (the logarithmic function). The generic factorable case is $\tilde{g}_{A,h}(\lambda,t) = f_{A,h}(\lambda) \cdot g_{A,h}(t)$. The special case written as generally as possible in case of multiple such forms is $\tilde{g}_{A,h}(\lambda,t) = \log(f_{A,h}(\lambda) \cdot g_{A,h}(t)) = \log(f_{A,h}(\lambda)) + \log(g_{A,h}(t))$ so that in (1) just below the λ-term absorbs into the normalization term and the *t*-term is just another of the added together (A)-functions with its $f_{C,h}(\lambda) = 1$. Each such (A)-function of *T* has an associated conditional expectation equal to a number, e.g. the generic case, $E[f_{A,h}(\Lambda=\lambda) \cdot g_{A,h}(T)|\lambda] = c_{A,h}$ where this number does not depend on λ, and for example in the special case, $E[\log(f_{A,h}(\Lambda=\lambda) \cdot g_{A,h}(T))|\lambda] = c_{A,h}$. The (B) functions lack a λ, e.g., $g_{B,h}(T)$, but the conditional expectation constraint varies with λ, e.g., $E[g_{B,h}(T)|\lambda] = c_{B,h}(\lambda)$. The (C) functions look like the (B) functions, $g_{C,h}(T)$, but have expected values that are λ-agnostic, e.g., $E[g_{C,h}(T)] = c_{C,h} \; \forall \; \lambda$, i.e., the constraint values are not explicitly a function of any one selected λ. (Equivalently and with appropriate constants $c_{subscript}$, the constraints appearing in the Lagrange equation before taking the derivative take three forms: (A) $\alpha_{A,h}(c_{A,h} - E[f_{A,h}(\Lambda=\lambda) \cdot g_{A,h}(T)|\lambda]) = 0$; (B) $\alpha_{B,h}(\lambda)(c_{B,h}(\lambda) - E[g_{B,h}(T)|\lambda]) = 0$; and (C) $\alpha_{C,h}(c_{C,h} - E[g_{C,h}(T)]) = 0$.) Finally, given that none of the constraining expectations are redundant or contradictory (see REMARKS B), then (i) $p(t|\lambda)$ is in the exponential family, i.e.,

$$p(t|\lambda) = \exp(-\alpha_0(\lambda) - 1 - \sum_h \alpha_{A,h} f_{A,h}(\lambda) \cdot g_{A,h}(t) - \sum_h \alpha_{B,h}(\lambda) g_{B,h}(t) - \sum_h \alpha_{C,h} p(\lambda) g_{C,h}(t)), \quad (1)$$

(ii) the set of g(*t*) functions is a sufficient statistic for $p(\lambda|t)$, and
(iii) the form of a minimal sufficient statistic does not change with sampling (for us a sample is a synaptic activation).

*Proof*: The constraint functions here are restricted to only those that factor functions of *t* from functions of λ or that are functions of *t* without λ appearing. Thus, via factorability [17], the distributions are restricted to the exponential family (recall that this is an if and only if condition). Because the constraints cannot change the range of *T*, this is the exponential family of fixed dimension. Item (iii) invokes the Koopman-Darmois theorem [18,19], which applies to the exponential family with distributions of fixed dimension (support). The term $\alpha_0(\lambda) - 1$ arises from the normalization constraint. The $p(\lambda)$ terms arise from taking the derivative of the unconditional expectations; i.e., $E[g_{C,h}(T)] = \int p(\lambda) \cdot E[g_{C,h}(T)|\lambda] d\lambda$ at any one conditioning value $\Lambda = \lambda$ while the constraint value is always λ-independent.

*Preliminary remarks*: Constraints that are merely a constant, such as the per-pulse cost of communication, do not appear explicitly as a term of (1) but exert their effects through the resolved value of the Lagrange multiplier(s). Functions such as $T^\Lambda, \exp(\Lambda T), \Gamma(\Lambda T)$, and $\tan(\Lambda T)$ do not factor Λ from *T*; thus as expectation constraints, e.g., $E[T^\Lambda]$, $E[\exp(T\Lambda)]$, etc., these are ruled out even though any one of them can be used as a constraint set to develop a maximum entropy distribution. Finally, if h(*T*), the differential entropy of *p*(*t*), is constant as inferred in model 3 below, then this inference is a minimum relative entropy statement as in [15].

We now have a clearer path for finding conditions for which *t* itself can be a sufficient statistic, i.e., conditions for which *t* carries all the Fisher information of *j*'s computational task. To move a little further down this path, and to get closer to some biology, we identify neural constraints that are, one-by-one, a representative of the (A)-, (B)-, and (C)-constraint types just defined.

IV. Constraints, Constraint Sets and their implied Likelihoods

*Energy constraints.*[1] (See Table 1) From a biological perspective and in the simple models considered here, energy constraints (per-pulse calculations) take a limited number of forms: (i) a constant per pulse cost, $\mathcal{E}_0$, overwhelmingly dominated by axonal communication costs, (ii) time-linear costs such as maintenance, axonal leak, and lifetime defrayed construction costs, (iii) synaptic activation costs, $\mathcal{E}_{syn}$, which are proportional to $\lambda \cdot t$, (iv) a time-logarithmic cost for re-setting timing-devices that tick with exponential spacing, and depending on the neuron model, (v) an inverse-time cost to account for discarded energy between IPIs. Here we choose to associate the time-linear costs with the decision-maker (DM) that controls the duration of a decision-making cycle (e.g., a fixation interval in the visual system or a theta cycle in the olfactory and whisking systems) in which *j* might generate many IPIs (such a DM must know these time-proportional costs to decide when additional system computation is not improving the decision's expected utility, which takes into account the energy-costs). Thus here, time-linear costs do not constrain *j*'s MEM and (vi), inverse-time costs, are also ignored.

TABLE 1  Notation for certain constants and constraint values

$c_1$     a scale constant with units of joules
$c_2$     another scale constant with units of joules
$c_{est}$   constant of proportionality when estimation is $\hat{\lambda} := c_{est}/t$
There are two equivalent constants when a logarithmic estimation relationship exists:
$c_{logest}$   estimation constant within logarithm
$c_{logest*}$   alternative estimation constant $c_{logest*} := -\log(c_{logest})$
$b_\lambda$   one second
$\mathcal{E}$   total energy-budget
$\mathcal{E}_0$   constant per pulse energy-costs
$\mathcal{E}_{syn}$   energy-cost of postsynaptic excitation
$\mathcal{E}_{log}$   energy-cost of clocking with exponentially separated ticks

In general the logarithmic energy constraint takes the form $c_2 E[\log(T/b_\lambda) | \lambda] = \mathcal{E}_{log}(\lambda)$, but there is an appealing $\mathcal{E}_{log}(\lambda)$ form that is motivated by the sensible inverse relationship between *t* and λ. That is, conjecture $\mathcal{E}_{log}(\lambda) := -c_2 \log(\lambda b_\lambda) + \mathcal{E}_{log} > 0$, which implies $c_2 E[\log(\lambda T) | \lambda] = \mathcal{E}_{log}$. In this case a B-type constraint becomes an A-type constraint.

In sum, the most complicated energy-constraint expression used here is
$\alpha_1(\mathcal{E}_{syn} - c_1 \lambda E[T | \lambda]) + \alpha_{log}(\mathcal{E}_{log} - c_2 E[\log(\lambda T) | \lambda])$ with energy budget $\mathcal{E} := \mathcal{E}_0 + \mathcal{E}_{syn} + E_\Lambda[\mathcal{E}_{log}(\Lambda)]$ or equivalently $\mathcal{E}_0 := \mathcal{E}_{syn} + \mathcal{E}_{log} - c_2 E_\Lambda[\log(\Lambda b_\lambda)]$, noting that there is a constant of per-pulse energy-use, $\mathcal{E}_0$, accounting for the cost of pulse-generation and communication incorporated into the total budget.

---

FOOTNOTE *1*: Because of the dimensional inhomogeneity of the constraints, care is taken with the constants needed for homogeneity and sensibility. Whenever λ and *t* (or *T*) appear, their product is dimensionless but whenever either appears alone, it is multiplied by an appropriate constant: in the case of rate λ (events/sec), it is multiplied by seconds, $b_\lambda$; and in the case of *t* (or *T*), it is divided by $b_\lambda$. The densities *p(t|λ)* only appear to have an unmatched λ outside the exponent because *dt* is left implicit.

---

*Estimation constraints.* Although the MEM corollary allows a very general set of possibilities including multiple estimates and estimates with λ-dependent bias, the constraints with intuitive appeal are more limited; in particular, $f_h(\lambda) = E[f_h(1/T) | \lambda] + c$. That is, estimations, where within the same function, λ appears on one side and 1/*T* appears on the other side; only two estimation forms occur below. Of these two, the simpler is the estimation of λ by a constant divided by $t$, $\hat{\lambda} := c_{est}/t$, and the less simple is estimation of the logarithm of λ, $\widetilde{\log(\lambda b_\lambda)} := \log(b_\lambda c_{logest}/t)$. The former, $\hat{\lambda}$, is constrained to be unbiased, $E[c_{est}/T | \lambda] = \lambda$. On the other hand, the latter is constrained to be biased with several equivalent forms, e.g., $E[\log(b_\lambda/T)|\lambda] = \log(\lambda b_\lambda) + c_{\log cest*}$, and with $c_{logest*} := -\log(c_{logest})$, $E[\log(b_\lambda c_{\log cest}/T)|\lambda] = \log(\lambda b_\lambda)$, *inter alia*.

*Unconditional estimation constraints.* The third summation of equation (1) arises when combining an estimation definition with an unconditional expectation of the latent variable. Here only one such constraint appears: From the estimation definition $\hat{\lambda} := c_{est}/t$ and given $E[\Lambda]$, then, through the integration across $p(\lambda)$ noted above, $E[c_{est}/T] = E[\Lambda]$, or as a constraint on $T$, $E[c_{est}/T] - E[\Lambda] = 0$.

A third and rather distinct type of λ-independent constraint can also be part of the constraint set. This constraint is fundamentally a communication constraint. This type of constraint controls the shape of the output-pulse waveform, which itself always will limit information rate at the level of intersymbol interference. For example, suppose that there is a random variable limiting how quickly one pulse can be generated after another, and suppose this constraint can be expressed as an expectation. For example, a constraint as $E[1/T]$ delays the rise of a pulse and has no dependency on λ.

The MEM Corollary is now applied to three specific constraint sets but 'without a fully defined neuron' (skeleton). To prove each result merely substitute the given constrants into equation (1). Each of these results is expressed in two ways since the normalizations are well-known.

*MEM Corollary Model 1 (skeleton).* Given constraint set $M1 := \{ \mu_w E[\Lambda T|\lambda], -E[\log(c_{logest}\Lambda T)|\lambda] \}$. Then, the likelihood of $\Lambda=\lambda$ takes the form of a Γ-distribution,
$$p(t|\lambda) = \exp(-\alpha_0(\lambda) - 1)(\lambda t)^{-\alpha_2} \exp(-\alpha_1 \mu_w \lambda t) \tag{2a}$$
$$p(t|\lambda) = (\alpha_1 \mu_w \lambda)^{-\alpha_2+1} t^{-\alpha_2} \exp(-\alpha_1 \mu_w \lambda t)/\Gamma(-\alpha_2+1) \tag{2b}$$
The first constraint is a synaptic energy-constraint with constant $\mu_w$, $\mathcal{E}_{syn} = \mu_w E[\lambda T|\lambda]$. The second constraint is a biased estimation constraint as already described.

*MEM Corollary Model 2 (skeleton).* Given constraint set $M2 := \{ \mu_w E[\Lambda T|\lambda], c_2 E[\log(\Lambda T)|\lambda], E[c_{est}/(\Lambda T)|\lambda] \}$. Then the distributional form is a generalized inverse Gaussian distribution (GIG) [20-22].
$$p(t|\lambda) = \exp(\alpha_0(\lambda) - 1)(\lambda t)^{-\alpha_3 - 1} \exp(-\alpha_1 \mu_w \lambda t - \alpha_2 c_{est}/(\lambda t)); \tag{3a}$$
$$p(t|\lambda) = (2K_{\alpha_3}(\sqrt{4\alpha_1 \mu_w \alpha_2 c_{est}}))^{-1}(\alpha_1 \mu_w \lambda^2/\alpha_2 c_{est})^{-\alpha_3/2} t^{-\alpha_3 - 1} \exp(-\alpha_1 \mu_w \lambda t - \alpha_2 c_{est}/(\lambda t)), \tag{3b}$$
where $K_{\alpha_3}(\cdot)$ is a modified Bessel function of the second kind with index value $\alpha_3$.

(In this and in the next model, the Lagrange multiplier of the logarithmic term is offset by one in anticipation of the GIG result and the standard form of this exponent in GIG distributions.) The first constraint is as in $M1$. The second constraint, where energy costs are always positive, arises as an energy-cost of clocking $c_2 E[\log(\lambda T)|\lambda] = \mathcal{E}_{\log}$ and, as already described, is a B-type constraint, which, as energy-use, must be positive. The third constraint is the unbiased estimation requirement with scale constant $c_{est}$, $E[c_{est}/T|\lambda] = \lambda$ with $j$'s λ estimate $\hat{\lambda} := c_{est}/t$.

> TABLE 2  Notation for *M3-skeleton* and implied marginals
>
> $S(\lambda)$    simplifying notation  $S(\lambda):=\alpha_2 c_{est}+\alpha_{2,1}\lambda p(\lambda)$
>
> $R$    simplifying notation  $R:=\alpha_2 c_{est}+\alpha_{2,1}c_\lambda$
>
> $\lambda_{min}$    lower bound ($> 0$) of the range of $\lambda$.
> $\lambda_{max}$    upper bound ($< \infty$) of the range of $\lambda$.
> $k$    ratio of upper to lower bound, $k := \lambda_{max}/\lambda_{min}$.
> $c_\lambda$    normalization term for $p(\lambda)$;  $c_\lambda := 1/log(\lambda_{max}/\lambda_{min})$

*MEM Corollary Model 3* (*skeleton*). (see Table 2) Given constraint set
$M3:=\{\ \mu_w E[\Lambda T|\lambda],\ c_2 E[\log(\Lambda T)|\lambda],\ E[c_{est}/(\Lambda T)|\lambda],\ E[\Lambda]\ \} \equiv \{\ \mu_w E[\Lambda T|\lambda],\ c_2 E[\log(\Lambda T)|\lambda],\ E[c_{est}/(\Lambda T)|\lambda],$
$E[1/T]\ \}$. Then the distributional form is again a generalized inverse Gaussian distribution.
$$p(t|\lambda) = \exp(\alpha_0(\lambda)-1)(\lambda t)^{-\alpha_3-1}\exp(-\alpha_1\mu_w\lambda t - \alpha_2 c_{est}/(\lambda t) - \alpha_{2,1}p(\lambda)/t)\ ; \text{ or} \tag{4a}$$
$$p(t|\lambda) = (2K_{\alpha_3}(2\sqrt{\alpha_1\mu_w S(\lambda)}\ ))^{-1}((\alpha_1\mu_w\lambda^2)/S(\lambda))^{-\alpha_3/2}t^{-\alpha_3-1}\exp(-\alpha_1\mu_w\lambda t - (\lambda t)^{-1}S(\lambda))\ , \tag{4b}$$
with  $S(\lambda):=\alpha_2 c_{est}+\alpha_{2,1}\lambda p(\lambda)$ .

The first three constraints are as in *M2*. The fourth constraint as utilized, is $E[1/T] = c_{est}^{-1}E[\Lambda]$. This constraint arises from knowledge of $E[\Lambda]$ and the unbiased constraint, $E[c_{est}/T|\lambda] = \lambda$ . That is, $E[T^{-1}] = \int p(\lambda)E[T^{-1}|\lambda]d\lambda = \int p(\lambda)c_{est}^{-1}\lambda d\lambda = c_{est}^{-1}E[\Lambda]$. Comparing (4b) to (3b), the presence of an apparent $\lambda$-dependency inside a Bessel function is notable in that it must be removed. See APPENDIX A for the proof of this statement and a further development. From this development, either $\alpha_{2,1} \propto 1/p(\lambda)$ or $p(\lambda) \propto 1/\lambda$ (each with its own constant). The following lemma uses this result and the assumption that $\alpha_{2,1}$ is not directly proportional to the inverse of $p(\lambda)$ across all values of the random variable. Motivating this assumption, is the fact that the unconditional constraint becomes redundant with the conditionally unbiased $\lambda$-dependent constraint when $\alpha_{2,1} \propto 1/p(\lambda)$.

*Lemma 1*. A sufficient condition for $S(\lambda)$ and for all the $\lambda$-conditional expectations of *M3* to be $p(\lambda)$-free is $p(\lambda) \propto \lambda^{-1}$. Additionally, a conjecture, supported by ½-integer $\alpha_3$ examples, is that this sufficient condition is also necessary.
*Proof*: Note that the terms of $\lambda$ in $S(\lambda)$ disappear when $p(\lambda) \propto \lambda^{-1}$. The '½-integer values of $\alpha_3$' examples, which always avoid Bessel functions, are generated by Mathematica calculating expectations.

*Lemma 2*. Assume that $p(\lambda)$ is a density, and from lemma 1, is proportional to $1/\lambda$, then
$p(\lambda) = c_\lambda/\lambda$ with $c_\lambda := (\log(\lambda_{max}/\lambda_{min}))^{-1}$ and $\lambda \in [0 < \lambda_{min} < \lambda_{max} < \infty]$.
*Proof*: A density integrates to one, and this particular density must be bounded away from zero and infinity for there to exist a (finite) normalization term.

From Lemma 2, the *MEM Corollary Model 3* (*skeleton*) simplifies:
*Lemma 3*:  $p(t|\lambda) = (2K_{\alpha_3}(2\sqrt{\alpha_1\mu_w R}\ ))^{-1}(\alpha_1\mu_w\lambda^2/R)^{-\alpha_3/2}\cdot t^{-\alpha_3-1}\exp(-\alpha_1\mu_w\lambda t - R/(\lambda t))$ \hfill (5)
with  $R:=\alpha_2 c_{est}+\alpha_{2,1}c_\lambda$  replacing $S(\lambda)$.

*Theorem defining the marginal distributions of M3*: Given constraint set *M3* and the MEM inferred $p(t|\lambda)$ for this constraint set,
$p(\lambda) = (\lambda\log(\lambda_{max}/\lambda_{min}))^{-1}$; \hfill (6)
defining $\lambda_{max}:=k\cdot\lambda_{min}$ and substituting into $E[\Lambda]=(\lambda_{max}-\lambda_{min})/\log(\lambda_{max}/\lambda_{min})$,

$\lambda_{min} = E[\Lambda]log(k)/(k-1)$ and $\lambda_{max} = kE[\Lambda]log(k)/(k-1)$ ; (7)
moreover,

$$p(t) = \int_{\lambda_{min}}^{\lambda_{max}} (\log(\lambda_{max}/\lambda_{min}) 2 K_{\alpha_3}(2\sqrt{\alpha_1 \mu_w R}))^{-1} (\alpha_1 \mu_w / R)^{-\alpha_3/2} \cdot (\lambda t)^{-\alpha_3 - 1} \exp(-\alpha_1 \mu_w \lambda t - R/(\lambda t)) d\lambda .$$ (8)

*Proof*: The first part of the theorem consists of the lemmata. Eq. (8) is the integral $\int p(\lambda) p(t|\lambda) d\lambda$ just over the appropriate interval, partially described by (7).

*Corollary $p(\lambda)$ 1*: Given the original constraint set and given the value of $\lambda_{max}$ or $E[\Lambda] \div \lambda_{max}$, $p(\lambda)$ is fully specified.
*Proof*: Recall $E[\Lambda]$ is in the constraint set so that combining (7) with either piece of additional knowledge is enough to solve for $k$.

*Example 1*: Use this corollary along with (6) and (7). Suppose 4000 inputs lines with an average firing rate of 8 Hz, then $E[\Lambda] = 32$ events/msec, and suppose $E[\Lambda]/\lambda_{max} = .02$, then $\lambda_{max} = 1,600$ events/msec and $\lambda_{min} \approx 3 \cdot 10^{-13} \approx 0$ events/msec; finally, $p(\lambda) = 0.02/\lambda$. The 0.02 reappears because when $\lambda_{max} \gg \lambda_{min}$, then $(\lambda_{max} - \lambda_{min})/\lambda_{max}$ is nearly one implying $E[\Lambda]/\lambda_{max}$ is nearly the normalization term $c_\lambda$. But this result is rather *ad hoc* and maybe inconsistent with the other constants and constraints.

*Corollary $p(\lambda)$ 2*: Combining $E[\Lambda]E[1/\Lambda]=(k-1)^2/(k \cdot (log(k))^2)$, (7), and the values of $E[\Lambda]$ and $E[1/\Lambda]$, where the former comes from the constraint set and the latter comes from the integral of the unbiased estimation constraint across $p(\lambda)$, then the marginal distributions are fully parameterized.

*Example 2*: In fact *Example 1* is problematic in its arbitrary supposition of $E[\Lambda]/\lambda_{max}$ because there is enough specified in the constraint set and the defined constants to solve for the limits of $\lambda$. Here a very different range is produced. For the present calculation suppose there are synaptic failures at a homogenous rate of 0.5, and thus a success rate $s=1/2$; let $N_{syn}=4000$, and the unfailed firing rates be $\gamma=8$Hz; then, $E[\Lambda_j]=4000 \cdot 8/2=16000$ events per sec. Moreover, $E[\Lambda_j]E[1/\Lambda_j]=4000/4\theta_\mu=1000/\theta_\mu$. Then $\theta_\mu$ must be less than 1000 for the required $k > 1$. If $\theta_\mu=500$, then $k \approx 19.75$; $\lambda_{min} \approx 2545$; and $\lambda_{max} \approx 50260$. Seen Appendix B for detailed calculations.

## V. EXISTENCE COROLLARIES THAT CREATE NEURAL RELEVANCE

Here we identify specific neural models that are consistent with the above results and that lead to relatively tractable versions of the foregoing results. The common assumption leading to tractability is a linearly additive integrate and fire neuron model with threshold $\theta$ that fully resets after each pulse-out. The following three developments are essentially existence proofs. The axioms *A0* through *A2* combined with one of the constraint sets, *M1*, *M2*, or *M3*, are each a nearly complete descriptions of a neuron. By adding a linear additivity assumption, some additional conclusions about the particular neuron, as a statistical processing device, follow. These insights include the unbiased (possibly asymptotic) nature of the neuron's estimation, ways to reduce the MSE of a neuron's estimate, and that the IPI *t* is a minimal sufficient statistic in each case. *Neuron-1* is matched with constraint set *M1* and *Neuron-2* is applied to both *M2* and *M3*. The matching consists of aligning and equating terms by their function of *t* (e.g., for the pair $exp(-at-b/t) = exp(-ct-d/t)$, then $a=c$ and $b=d$). In all cases $\Lambda$ is assumed to be a construction of a union of point processes for which the Poisson approximation [23] is valid.

TABLE 3  Notation for *neuron-1*

$\theta$      threshold to spike (a unitless integer)
$\mu_w$      energy scale constant of a synaptic event
$c_{logest}$      estimation constant within logarithm
$c_{logest*}$      alternative estimation constant $c_{logest*} := -log(c_{logest})$

*NEURON-1 DEFINED*: (see Table 3) In addition to the axioms and *M*1, all the synapses are excitatory and of equal value, $\mu_w$. There is no inhibition or leak; that is, the neuron just adds synaptic excitatory synaptic events. The conditional distribution is the well-known waiting-time result [5, 24, 25] with unit steps and a mean Poisson rate of $\lambda$. Specifically with the integer threshold $\theta \geq 1$,

$$p(t|\lambda) = \lambda^\theta t^{\theta-1} \exp(-\lambda t)/\Gamma(\theta). \tag{9}$$

Furthermore, recall that *M1* is performing the estimate
$\widehat{\log(b_\lambda \lambda)} := \log(b_\lambda c_{\log cest}/T) = \log(b_\lambda/T) - \log(1/c_{\log cest*})$ that is biased as
$E[\log(b_\lambda/T)|\lambda] = \log(b_\lambda \lambda) + \log(1/c_{\log cest*}) = \log(b_\lambda \lambda/c_{\log cest*})$. Therefore, the MSE(*M1*):=
$E[(\log(b_\lambda c_{\log est}/T) - \log(b_\lambda \lambda))^2 | \lambda] = c^2_{\log est} E[\log^2(b_\lambda/T)|\lambda] - E[\log(b_\lambda/T)|\lambda]^2$.

*MEM Corollary M1, Neuron-1*. Given *neuron-1*, constraints *M1*, and noting that the inferred distribution of *M1-skeleton* aligns with *neuron-1*'s FHT distribution, then

(i) Because $\mathcal{E} = \mathcal{E}_{syn} + \mathcal{E}_0 = \theta\mu_w + \mathcal{E}_0$, $\theta = (\mathcal{E}-\mathcal{E}_0)/\mu_w$. Then aligning terms of (2b) and (9) implies $\alpha_1 \mu_w = 1$, and $-\alpha_2 = (\mathcal{E}-\mathcal{E}_0)/\mu_w - 1$. To resolve the bias term: from just above, $E[\log(\lambda T)|\lambda] = \log(c_{\log est})$, and taking the expectation, $c_{\log est} = Di\Gamma(\theta) = Di\Gamma((\mathcal{E}-\mathcal{E}_0)/\mu_w)$.
(ii) *t* is a minimal sufficient statistic for the synaptic activation based estimate of $\Lambda$'s value;
(iii) the MSE(*M1*) of the logarithmic estimate is
$E[(\log(T/b_t) - E[\log(T/b_t)|\lambda])^2 | \lambda] = Tri\Gamma[(\mathcal{E} - \mathcal{E}_0)/\mu_w] \approx ((\mathcal{E} - \mathcal{E}_0)/\mu_w)^{-1}$, asymptotically unbiased as threshold and energy-budget increase without limit.

---

TABLE 4 Notation for *neuron-2-divisive*

$\theta$      threshold to spike (energy on membrane capacitance)
*I*      divisive inhibitory constant
$\mu_+$      energy of average excitatory events, $E[W_+]$
$E[W_+^2]$  the second-moment of excitatory events
$\lambda$      rate of excitatory synaptic arrivals
*U*      simplifying notation, $U := \mu_w/I$
*Q*      simplifying notation, $Q := E[W_+^2]/I^2$

---

*NEURON-2 DIVISIVE INHIBITION DEFINED*: (see Table 4) Our primary exhibits are a little more biological than *Neuron-1* since they allow for synaptic inhibition and differently valued synaptic weights. The threshold for this neuron is $\theta > 0$, and here we define threshold in units of energy (voltage-squared on the membrane capacitor). We need to assume that synaptic events, valued $w_{ij}$ (and once again energy), are additive and small but numerous to invoke the Gerstein and Mandelbraut approximation of drifted Brownian motion [26], and we need to assume that the Poisson approximation is valid when applied to the union of all excitatory synaptic events conditional on $\lambda$. In which case there is approximate $\lambda$-conditional independence. Inhibition is assumed to be divisive and a unitless constant, *I*. Both examples are satisfied by a Poisson process. Because *neuron-2-Divisive* does not allow interactions of membrane potential and threshold voltage, (i.e., synaptic events are linearly additive throughout the relevant voltage-range), the GIG family of densities collapses to the inverse Gaussian (IG) [27] itself (see also [23, 28]). Compared to *neuron-1*, the relation of this neuron's synaptic excitation needs to be expanded to include the average charge (energy)-injection by an active synapse, $\mu_+ := E[W_+]$ so that the average drift rate is, $\lambda\mu_+/I$, or at FHT, $\theta = \lambda\mu_+ E[T|\lambda]/I$. The associated variance is $\lambda E[W_+^2]/I^2$ (as long as both the variation of synaptic event-arrivals overwhelms Na-channel shot-noise and thermal noise see [29]); i.e., the diffusion constant is $\sqrt{\lambda E[W_+^2]}/I$. Writing the average synaptic weight downgraded by inhibition as $U := E[W_+]/I$ and defining its associated second-moment as $Q := E[W_+^2]/I^2$, the FHT distribution is essentially the Gerstein and Mandelbrot [26] IG result,

$$p(t|\lambda) = \theta(\pi Q\lambda)^{-1/2} \exp(2\theta U/Q) t^{-3/2} \exp(-U^2\lambda t/Q - \theta^2/Q\lambda t) \qquad (10)$$

From (10) the following useful moments are easily calculated with Mathematica:

$E[T|\lambda] = \theta/\lambda U = \theta I/\lambda\mu_+$; $E[1/T|\lambda] = \lambda(Q/2\theta^2 + U/\theta) = \lambda(Q+2U\theta)/2\theta^2 = \lambda(E[W_+^2]+E[W]\theta I)/2\theta^2 I^2$;

and, recalling that for $M2$ and $M3$ the unbiased estimate, $\hat{\lambda} := c_{est}/t$, is the constraint $\lambda = E[c_{est}/T|\lambda]$, then MSE($M2$ or $M3|\lambda$) := $E[((c_{est}T^{-1})-\lambda)^2|\lambda] = E[(c_{est}/T)^2|\lambda] - E[c_{est}/T|\lambda]^2 =$

$c_{est}^2 Q^2\lambda^2(1+U\theta/Q)/2 = c_{est}^2\lambda^2 E[W_+^2]^2(1+\mu_+\theta I/E[W_+^2])/2(\theta I)^4$.

*MEM Corollary M2, Neuron-2-Divisive.* Given *neuron-2-Divisive*, constraints $M2$, and noting that the inferred distribution of *M2-skeleton* aligns with *neuron-2*'s FHT distribution (10), then for (3b)

(i) $\alpha_3 = 1/2$; thus allowing the reduction

$\exp(\alpha_0(\lambda-1)) = (2K_{\alpha_3}(\sqrt{4\alpha_1\mu_+\alpha_2 c_{est}}))^{-1}(\alpha_1\mu_+\lambda^2/\alpha_2 c_{est})^{-\alpha_3/2} = \theta(\pi Q\lambda)^{-1/2}\exp(2U\theta/Q)$. For the terms in

the exponent of (3b), $\alpha_1\mu_+ = U^2/Q$; and $\alpha_2 c_{est} = \theta^2/Q$. Because $\mathcal{E} = \mathcal{E}_{syn} + \mathcal{E}_{log} + \mathcal{E}_0 = \theta \cdot I + \mathcal{E}_{log} + \mathcal{E}_0$,

$\theta \cdot I = (\mathcal{E} - \mathcal{E}_{log} - \mathcal{E}_0)$. Indeed, whenever the product $\theta \cdot I$ appears, it can be read as the energy needed to reach threshold. Moreover, the average number of events to reach threshold is

$E[\lambda T|\lambda] = \sqrt{(\theta^2/Q)/(U^2/Q)} = \theta/U = \theta I/\mu_+$.

(ii) $t$ is a minimal sufficient statistic (see REMARKS C);

(iii) $c_{est}$, the constant of the estimation is derived from the unbiased statement, $E[c_{est}/T] = \lambda$, and

$E[1/T|\lambda]$ as above to yield $c_{est}^{-1} = (Q+2U\theta)/2\theta^2 = (E[W_+^2]+2\mu_+\theta I)/2\theta^2 I^2$, or

$c_{est} = 2\theta^2 I^2/(E[W_+^2]+2\mu_+\theta I)$, or $c_{est}^2 = 4\theta^4 I^4/(E[W_+^2]+2\mu_+\theta I)^2$; and finally, (iv)

MSE($M2|\lambda$) $== c_{est}^2\lambda^2 E[W_+^2]^2(1+\mu_+\theta I/E[W_+^2])/2(\theta I)^4 = 2\lambda^2(1+\mu_+\theta I/E[W_+^2])/(1+2\mu_+\theta I/E[W_+^2])^2$ (11)

Thus, the MSE of the estimate decreases, more or less, in proportion to the synaptic energy-budget,

$(\theta \cdot I)^{-1} = (\mathcal{E} - \mathcal{E}_{log} - \mathcal{E}_0)^{-1}$; i.e., suppose $\mu_+\theta \cdot I/E[W_+^2] \gg 1$, then MSE *ca.* equals

$2\lambda^2(\mu_+\theta I/E[W_+^2])/(2\mu_+\theta I/E[W_+^2])^2 = \lambda^2 E[W_+^2]/(2\mu_+\theta I)$.

In sum, all the constraint variables and the neuron's estimation performance are now interrelated using the physical attributes of the defined neuron. For comments on the possible biophysics of a GIG FHT, see REMARKS D. For the next neuron we skip over some of these energy/estimation relationships to emphasize the special, inferred constraints on the system controlling the input latent variable and to emphasize the difference between the local (λ-dependent) and long-term (λ-independent) IPI probability distributions.

*MEM Corollary M3: Neuron-2.* Given *neuron-2*, constraints $M3$, noting that the inferred distribution of *M3-skeleton* aligns with *neuron-2*'s FHT distribution when taking advantage of *lemma* 3, then for (5)

(i) $\alpha_3 = 1/2$; thus allowing the reduction

$\exp(\alpha_0(\lambda-1)) = (2K_{\alpha_3}(2\sqrt{\alpha_1\mu_w R}))^{-1}(\alpha_1\mu_w\lambda^2/R)^{-\alpha_3/2} = \theta(\pi Q\lambda)^{-1/2}\exp(2U\theta/Q)$; $\alpha_1\mu_w = U^2/Q$; and

$\alpha_2 c_{est} + \alpha_{2,1} c_\lambda = R = \theta^2/Q = \theta^2 I^2/E[W_+^2]$. That is,

$$p(t|\lambda) = \theta(\pi Q\lambda)^{-1/2}\exp(2U\theta/Q) \cdot t^{-3/2}\exp(-U^2\lambda t/Q - \Theta^2/(Q\lambda t)) \qquad (12)$$

(ii) is unchanged from the previous example;

moreover, using (8) and (i),

(iii) $p(t) = (t\log(\lambda_{max}/\lambda_{min}))^{-1}(\Psi(A_{max}(t)) - \Psi(A_{min}(t)) - \exp(4U\theta/Q)(\Psi(B_{max}(t)) - \Psi(B_{min}(t))))$ (13)

with $\Psi(x) := (erf(x)+1)/2$; $A_h(t) := U\sqrt{t\lambda_h/Q} - \theta/\sqrt{tQ\lambda_h}$; $B_h(t) := U\sqrt{t\lambda_h/Q} + \theta/\sqrt{tQ\lambda_h}$; with index $h \in \{min, max\}$.

Note that the conditional distribution is in the exponential family; the λ-marginal distribution is in the exponential family; but the *t*'s marginal distribution is not in this family. The next section returns to an earlier interest (e.g., [in communication 46; in physics 11, 47, 48, 49; and in neuroscience 1-5, 30, 34]) in maximizing the bits/J. For another form of the λ- and *t*-marginals that develop not by using MEM but by making certain minimal assumptions, see [5].

## VI. THE NEURAL MODEL WITH OPTIMAL ESTIMATION FIXES THE BITS/J

At this point using *M3* and *Neuron-2* – with both a likelihood and a marginal inferred from the constraint set – there is now a way to return to the older problem of describing a neuron as an optimal, bits/J device. Note that joules-per-pulse is exactly the given energy-budget, e.g., $\mathcal{E} = \mathcal{E}_o + \mathcal{E}_{syn} + \mathcal{E}_{\log}$. Given the availability of the joules-per-pulse cost, we only need bits-per-pulse to get bits/J (although the calculations below treat information as nats).

To obtain a mutual information [31] while remaining in the context of computation for the purpose of estimating Λ's value, take the viewpoint of any recipient of *j*'s output. Then equip this recipient with the appropriate prior distribution and apply Lindley's [32] insight: a Bayesian experiment is quantifiable just as the noisy channel problem. That is, the Bayes inference is $p(\lambda|t) = \frac{p(t|\lambda)p(\lambda)}{\int_\lambda p(t|\lambda)p(\lambda)d\lambda} = \frac{p(t|\lambda)p(\lambda)}{p(t)}$

(which also happens to be in the exponential family) allows calculation of the mutual information, which can be written in either of two ways, $E_{\Lambda,T}[\log \frac{p(\Lambda|T)}{p(\Lambda)}] = E_{\Lambda,T}[\log \frac{p(T|\Lambda)}{p(T)}]$. Thus from any recipient's perspective, an IPI (*T=t*) is the outcome of an "experiment" that advances a recipient's prior. The final outcome of this metaphorical experiment, which produced conditionally independent samplings that are the arrival-times of the synaptic input events, is the sufficient statistic *t*. For *M3-skeleton* and for *M3-neuron-2*, the specific forms of these mutual informations are found in APPENDIX C. Of course a single neuron in this feedforward mode *never* does any of these calculations, but a biologist seeking evidence for optimization, or an engineer wanting to build energy-optimized neuron for estimation and information transmission, might usefully perform these computations.

## VII. SUMMARY AND CONCLUSIONS

The MEM as introduced here extends to other neural computations including discrimination and prediction and to more complicated neurons with individual dendritically localized inferences, and with some work, may also extend to network-level analyses. Certainly an energy-efficient network computation is built from individual energy-efficient neuron computations.

A comparison with Friston's free-energy method [33] and its maximum entropy method application [34, particularly pages 9-10] shows how different the results can be using the same method but with very different definitions of the random process and random variable of interest and as well, the particular constraint set. Indeed, our results emphasize the critical nature of the constraint set (see REMARKS B). Of course, the great difference between the results here and the results of Friston and colleagues is not too surprising. It seems their analysis, quite boldly, concentrates on the neural system. Here the goal is more modest; we emphasize random variables and constraint sets of particular relevance to microscopically testable predictions of optimal neuron function (see REMARKS B, and just below).

Returning again to the two distinct perspectives of an engineer vs. a biologist, it is clear that the results here tell the engineer the minimum energy-costs of computation (where computation has two meanings, estimation and Bayesian inference). For the biologists there are experimental predictions as in [4,35]. That is, Nature has already done the experiment by producing differently parameterized neurons across species and brain regions, and the results here predict how one measurable variable changes as a function of another variable. In the earlier work the variable pairs were relatively simple, energy-use vs. firing-rate and quantal-failure-rate vs. firing-rate; here the variables are the variables of various energy-uses vs. the constraint parameters, e.g., $c_\lambda$ or $\lambda_{max}$. As noted in the remarks below, a constraint set $\{f_h(\lambda) = E[g_h(T)|\lambda]; h \in \{1,...,m\}\}$ implies the equalities $\{\lambda = f_h^{-1}(E[g_h(T)|\lambda]) \ \forall h\}$ when the $f_h$'s are invertible, as they are here (additionally, unconditional constraining expectation statements imply conditional statements after taking the derivative or, as in the case of the axonal constraint $E[1/T]=c_\lambda$, must hold exactly for every λ since the axon is agnostic regarding the latent variable and its estimation). Through the λ-linked equalities and the inferred likelihood, every parameter (e.g., $c_\lambda$, $c_{est}$) can be expressed as a function of energy and expectations. Thus not only are IPI distributions predicted for experimental confirmation/falsification (note that the predicted λ-conditional distributions are GIG but the predicted unconditional IPI distribution is not) but so are the matchings between constraint parameters and measureable energy-use. Indeed the small examples given at the end of section IV are encouragement for inter-relating synapse counts and firing rates as experimental predictions.

## VII. REMARKS
*A.     History and motivations for using the MEM:*

Justification of the maximum entropy method as a way to discover the best density for inference has a lengthy history with the essential, pioneering work due to Jaynes [13, 14]. Of four justifications claiming that the MEM finds the best density, perhaps the one with strongest practical appeal is game-theoretic. (Three others are (i) axiomatic derivations [15, 36-39], (ii) the entropy concentration of possible outcomes in which the best density is the one that is consistent with an exponentially vast number of the outcomes [40,41], and (iii) the Jaynes' perspective, use all the constraints that are credible but do not contradict oneself by using a distribution requiring additional constraints that one believes not to exist [42], i.e., logical consistency is superior to logical inconsistency). Via game theory, there are two salutary effects of a MEM inference: (i) this density maximizes the long-term, average pay-off while (ii) minimizing the maximum chance of going broke [43].

*B.     Non-redundant , non-contradictory constraints*

A hallmark of the present treatment is the simultaneous application of multiple constraints, each of which may be fundamentally different (energy, estimation, and normalization here, but action potential waveform constraints can also be applied). In contrast to textbook MEM- applications [44], here parametric as well as logical consistency between constraints is an important issue. Care should be taken to avoid contradictions among constraints. A variant of *M1-neuron-1* with a two term energy-constraint, $c_1 E[\lambda T | \lambda] + c_2 E[\log(T) | \lambda] = \mathcal{E}_0$, creates the possibility of a contradiction due to the logarithmic energy-constraint. Because energy is positive, $c_2$ must be greater than zero, but then the term $t^{-\alpha_1}$ is worrisome because if the constraint forces $\alpha_1 \geq 1$, then normalization is impossible. A more obvious contradiction is the constraint set $\{E[1/T] = c_1, E[1/T] = c_2 : c_1 \neq c_2\}$. However, the set $\{E[1/T] = c_1, E[1/T | \lambda] = c_2 \lambda\}$ can be allowed but only if $c_1 = \int_\lambda c_2 \lambda \cdot p(\lambda) d\lambda = c_2 E[\Lambda]$. Another obvious acceptable condition, requiring $c_1 \leq c_2$ is $\{E[1/T] = c_1, E[1/T] \leq c_2\}$, just a single, non-contradictory constraint. Here is another way that a single expectation can show up twice without contradiction (assuming a sufficient energy-budget, as is implicit in all examples): $E[\log(c_{est}/T) | \lambda] = \lambda$ occurs as an estimation constraint, and there is a two-term energy-constraint $E[c_2 \log(T) | \lambda] + c_1 E[\lambda T | \lambda] = \mathcal{E}_0(\lambda)$ where the λ-dependency of the energy available (consistent with the integral noted previously) seems necessary when two costs, running at different rates (here linear vs. logarithmic), are linked together through a single constraint.

Finally, there is an unavoidable parametric consistency issue that, at the same time, allows one to see the relationship of every constraint parameter to the energy-budget. The constraint set $\{f_h(\lambda) = E[g_h(T)|\lambda]; h \in \{1,...,m\}\}$ implies the equalities $\{\lambda = f_h^{-1}(E[g_h(T)|\lambda]) \ \forall h\}$ when the $f_h$'s are invertible, as they are here (additionally, unconditional constraining expectation statements imply

conditional statements after taking the derivative or, as in the case of the axonal constraint $E[1/T]=c_\lambda$, must hold exactly for every λ since the axon is agonistic regarding the latent variable and its estimation). Through λ, this set of equalities interlinks the constants associated with all the individual constraining expectation-statements. As illustrated in the examples, the undetermined multiplier calculations link the energy-budget to the estimation constants. Such parametric relationships between the constants are design requirements for an engineer and are experimental objects to be measured for the neuroscientist. That is, *the values of the interrelated constraint parameters are predictions of the theory subject to measurement, falsification, and adjustment by empirical neuroscience*.

*C.    The sufficient statistic is a scalar*

Each of these three models reduce *j*'s computation and communication problem to the desired scalar variable *T*, even though the optimal distributional form of *M2* and *M3* are nominally a two (or three variable from a certain viewpoint) sufficient statistic. This reduction to a scalar arises from the Poisson approximation. However, because the additive independent increment diffusion is one parameter (the variance can be inferred from the mean drift rate), the sufficient statistic of both the diffusion and the FHT-distribution is a scalar. Communication with such a scalar sufficient statistic *t* is possible so long as a recipient has knowledge of θ. In the same vein, Bayesian inference by the recipient, using *p*(*t*|λ) and equipped with a prior *p*(λ) arising from the previous IPI, is then also possible.

*D.    GIG distributions: a biophysical speculation*

In the Gerstein and Mandelbrot model [26], the primary assumptions are a high frequency Poisson process, very small synaptic events, and the additivity of these events: because these events are small, numerous, and independent, their sum approximates a drifted Brownian motion (B-M). The jump-over threshold issue for non-infinitesimal events can never occur, physically, because actual synaptic charge-injections are smoothly rising with finite derivatives. The critical assumptions for approximating B-M with such small events is (λ-conditional) independence and additivity of the synaptic events. Indeed and in contrast to passive neurons which are subadditive and lead to an Ornstein-Uhlenbeck (O-U) process [43] (and therefore fail the requirement for a finite number of moment constraints for the FHT distribution [44]), neurons with active conductances can be additive linear in time and voltage. The Skellam and Poisson processes of *neuron-2* have the critical property of independence and an active-channel neuron can supply additivity, but the question can be asked: How do GIG-distributions arise as opposed to just the IG itself? First, recall that it is one or more logarithmic constraints that give rise to the term $t^{a-1}$, where the exponent *a* determines the specific member of the GIG family. Moreover, regardless of this exponent's value (< 0), every GIG is a FHT outcome [23], i.e., the solution to the Laplace-transformed Chapman-Kolmogorov continuous, backward-equation. However, different family-members occur because of a positional nonstationarity even while temporally stationarity holds. For example, the background, macroscopic process (e.g., a Poisson or Skellam) is temporally stationary, but the microscopic parameters need not be positionally stationary. As the barrier (*θ*) is approached, the microscopic parameters (drift and diffusion) of the backward-equation change their values. When $a = -½$, there is no barrier interaction; when $a > -½$, then the barrier exerts a distance-dependent repulsion; and when $a < -½$, the distance-dependent force is attractive. That is, the amplitude, but not frequency, of the synaptic events are modulated, much as could happen at the action-potential initiation-site where voltage-sensitive, Na- and K-channels activate as threshold is approached.

VIII. APPENDIX

A. That the presence of an apparent λ-dependency inside the Bessel function of (4b) must be removed follows as:

Given (4b), then $E[T^{-1}|\lambda] = \dfrac{\lambda\sqrt{\alpha_1\mu_w}K_{\alpha_3-1}(2\sqrt{\alpha_1\mu_w S(\lambda)})}{\sqrt{S(\lambda)}\cdot K_{\alpha_3}(2\sqrt{\alpha_1\mu_w S(\lambda)})}$. But from the defined constraint of

unbiasedness, $E[c_{est}/T|\lambda] = \lambda$, $c_{est}^{-1} = \dfrac{\sqrt{\alpha_1\mu_w}K_{\alpha_3-1}(2\sqrt{\alpha_1\mu_w S(\lambda)})}{\sqrt{S(\lambda)}\cdot K_{\alpha_3}(2\sqrt{\alpha_1\mu_w S(\lambda)})}$; however, $c_{est}^{-1}$ has no λ-dependence. A similar situation exists for the conditional mean, $\frac{\mathcal{E}_{syn}}{\lambda\mu_w} = E[T|\lambda] = \frac{\sqrt{S(\lambda)}K_{\alpha_3+1}(2\sqrt{\alpha_1\mu_w S(\lambda)})}{\lambda\sqrt{\alpha_1\mu_w}\cdot K_{\alpha_3}(2\sqrt{\alpha_1\mu_w S(\lambda)})}$ ; thus,

$\frac{\mathcal{E}_{syn}}{\mu_w} = \frac{\sqrt{S(\lambda)}K_{\alpha_3+1}(2\sqrt{\alpha_1\mu_w S(\lambda)})}{\sqrt{\alpha_1\mu_w}\cdot K_{\alpha_3}(2\sqrt{\alpha_1\mu_w S(\lambda)})}$, and again, by its definition, the LHS has no λ-dependence. Noting the relationship between Bessel functions of the second kind with successive indices, $K_{a+1}(x) - K_{a-1}(x) = K_a(x)\cdot 2a/x$ and after a couple of scalings, subtract these two λ-independent equalities to get $\mathcal{E}_{syn}\alpha_1 - c_{est}^{-1}S(\lambda) = \frac{\sqrt{S(\lambda)\alpha_1\mu_w}K_{\alpha_3+1}(2\sqrt{\alpha_1\mu_w S(\lambda)})}{K_{\alpha_3}(2\sqrt{\alpha_1\mu_w S(\lambda)})} - \frac{\sqrt{S(\lambda)\alpha_1\mu_w}K_{\alpha_3-1}(2\sqrt{\alpha_1\mu_w S(\lambda)})}{K_{\alpha_3}(2\sqrt{\alpha_1\mu_w S(\lambda)})}$

$= \frac{\sqrt{S(\lambda)\alpha_1\mu_w}}{K_{\alpha_3}(2\sqrt{\alpha_1\mu_w S(\lambda)})}(K_{\alpha_3+1}(2\sqrt{\alpha_1\mu_w S(\lambda)}) - K_{\alpha_3-1}(2\sqrt{\alpha_1\mu_w S(\lambda)})) = \sqrt{S(\lambda)\alpha_1\mu_w}(2\alpha_3/2\sqrt{\alpha_1\mu_w S(\lambda)}) = \alpha_3$.

In this development, the expectations were calculated with Mathematica. The sequential relationship of these particular Bessel functions is given at http://www.efunda.com/math/bessel/modifiedbessel.cfm.

Rearranging, $-c_{est}^{-1}S(\lambda) = \alpha_3 - \mathcal{E}_{syn}\alpha_1$; $S(\lambda) = c_{est}(\mathcal{E}_{syn}\alpha_1 - \alpha_3)$, or $\alpha_{2,1}\lambda p(\lambda) = c_{est}(\mathcal{E}_{syn}\alpha_1 - \alpha_3 - \alpha_2)$, but none of the terms on the RHS have any dependence on $p(\lambda)$. Thus either $\alpha_{2,1} \propto 1/p(\lambda)$ or $p(\lambda) \propto 1/\lambda$ (each with its own constant).

B. *EXAMPLE 2*. For *M3*, the Theorem of the marginal distributions establishes (6), $p(\lambda) = (\lambda\log(\lambda_{max}/\lambda_{min}))^{-1} = (\lambda\log(k))^{-1}$. This section shows how to calculate the parameters of this distribution, i.e., any two of $\{k,\lambda_{max},\lambda_{min}\}$ from the relevant biological data. After calculating the required expectations by exact integration, one obtains the useful relationship $E[\Lambda_j]E[1/\Lambda_j] = (k-1)^2/(k^{1/2}\log(k))^2$. The numerical value of this equality will now be expressed in terms of biological variables and the relevant, given constraint values.

Let all synapses be valued one. Recall that $\lambda_{inh} = \Delta\cdot\lambda_{ex}$ allowing a multiplicative inhibition that decreases excitation by the factor $(1-\Delta)$. Whatever the synaptic energy factor $\mu_w$ is valued, it will be convenient to absorb it into the threshold, $\theta_\mu := \theta/\mu_w$, which allows all the calculations that follow to be unitless because threshold θ is in joules as is $\mu_w$. Given a particular failure rate of synaptic transmission, the success fraction designated, *s*, is the complement of this failure rate. Recall $E[\Lambda_j]$ is a constraint in *M3*, and thus its value is known. The latent RV $\Lambda_j$ is constructed by a weighted sum of the individual point process rates; then for the overall mean rates $E[\Lambda_j] = \sum_i 1\cdot E[\Lambda_i]$. All the Λ RVs take into account synaptic failures. Thus with $\gamma_i s := E[\Lambda_i]$ for all *i*, $\gamma_i$ is the long-term average output firing rate of the $i^{th}$ input to *j*. As a further simplification for this example, suppose all average firing rates are the same, including *j*'s output rate. This last assumption allows us to write two important statements. The first replaces all the $E[\Lambda_i]$'s in the summation;
$E[\Lambda_j] = N_{syn}\gamma s$.  (A1)

The second equality arising from the statement of equivalent output rates is
$\gamma = 1/E[T_j]$; therefore substituting into (A1),
$E[\Lambda_j] = sN_{syn}/E[T_j]$  (A2).
But consider the conditional average needed to reach threshold;
$(1-\Delta)E[\lambda T_j|\lambda] = \theta_\mu$. From this expression we can obtain the unconditional expectation of the inverse of $\Lambda_j$.
$E[T_j|\lambda] = \theta_\mu((1-\Delta))^{-1}/\lambda$; implying $E[T_j] = \theta_\mu((1-\Delta))^{-1}E[1/\Lambda_j]$.
Substituting into A1, $E[\Lambda_j] = sN_{syn}(1-\Delta)/(\theta_\mu E[1/\Lambda_j])$, which leads to the sought relationship,
$E[\Lambda_j]E[1/\Lambda_j] = sN_{syn}((1-\Delta)\mu)/\theta_\mu$

For the present calculation suppose $N_{syn} = 4000$, $s = 1/2$ and $\gamma = 8$Hz; then, $E[\Lambda_j] = 4000\cdot 8/2 = 16000$ events per sec. Then, $E[\Lambda_j]E[1/\Lambda_j] = 4000/4\theta_\mu = 1000/\theta_\mu$. Then $\theta_\mu$ must be less than 1000 for the required $k > 1$. If $\theta_\mu = 500$, then $k \approx 19.75$; $\lambda_{min} \approx 2545$; and $\lambda_{max} \approx 50260$.

C. Lindley-Shannon mutual information and calculations.

In the case of *M3-neuron-2*, to calculate mutual information one uses (12) and (13), first writing the mutual information as the difference of differential entropies, $E_{\Lambda,T}[\log \frac{p(T|\Lambda)}{p(T)}] = h(T) - h(T|\Lambda)$. Not much can be done with the marginal entropy,
$h(T) = E_T[\log(T)] + \log(\log(\lambda_{max}/\lambda_{min}))$
$- E_T[\log(\Psi(A_{max}(T)) - \Psi(A_{min}(T)) - \exp(4U\theta/Q)(\Psi(B_{max}(T)) - \Psi(B_{min}(T))))]$
with $E_T[\log(T)] = -\log(U\sqrt{\lambda_{max}\lambda_{min}}/\theta) - 2(\sqrt{U\theta}\exp(2U\theta/Q)K_{1/2}^{(1,0)}(2U\theta/Q)/(\sqrt{\pi Q})$
But for the conditional entropy some reduction is possible
$h(T|\Lambda) = \log(\pi Q\theta/(U^3\lambda_{max}\lambda_{min}))/2 - 3\sqrt{U\theta/\pi Q} \cdot \exp(2U\theta/Q) \cdot K_{1/2}^{(1,0)}(2U\theta/Q)$. Note that because we know the marginal in λ, we can calculate
$E_T[\log(T)] = \log(\theta/U) - \log(\lambda_{max}\lambda_{min})/2 - 2\sqrt{U\theta/\pi Q} \cdot \exp(2U\theta/Q) \cdot K_{1/2}^{(1,0)}(2U\theta/Q)$. Then subtracting the conditional from the marginal entropy yields
$E_{\Lambda,T}[\log \frac{p(T|\Lambda)}{p(T)}] = (-1 + \log(U\theta/\pi Q))/2 + \log(\log(\lambda_{max}/\lambda_{min}))$
$- E_T[\log(\Psi(A_{max}(T)) - \Psi(A_{min}(T)) - \exp(4U\theta/Q)(\Psi(B_{max}(T)) - \Psi(B_{min}(T))))]$
$+ 3\sqrt{U\theta/\pi Q} \cdot \exp(2U\theta/Q) \cdot K_{1/2}^{(1,0)}(2U\theta/Q)$
Directing our attention to the term $\log(U\theta/\pi Q))/2$, we see from the definition of U and Q that information is growing logarithmically with linear increases of threshold or inhibition, or the ratio $\mu_+/E[W_+^2]$.